\documentclass[%
 preprint,
 amsmath,amssymb,
 aps,
]{revtex4-2}

\usepackage{xcolor}
\usepackage{graphicx}
\usepackage{dcolumn}
\usepackage{bm}
\usepackage[italicdiff]{physics}
\usepackage{hyperref}

\begin{document}


\title{Effects of bursty synthesis in organelle biogenesis}

\author{Binayak Banerjee$^1$}
 
\author{Dipjyoti Das$^1$}%
 \email{dipjyoti.das@iiserkol.ac.in}
\affiliation{%
$^1$Department of Biological Sciences, Indian Institute of Science Education And Research Kolkata, 
Mohanpur, Nadia - 741 246,
West Bengal, India}%

\begin{abstract}

 A fundamental question of cell biology is how cells control the number of organelles. The processes of organelle biogenesis, namely \textit {de novo} synthesis, fission, fusion, and decay, are inherently stochastic, producing cell-to-cell variability in organelle abundance. In addition, experiments suggest that the synthesis of some organelles can be bursty. We thus ask how bursty synthesis impacts intracellular organelle number distribution. We develop an organelle biogenesis model with bursty \textit{de novo} synthesis by considering geometrically distributed burst sizes. We analytically solve the model in biologically relevant limits and provide exact expressions for the steady-state organelle number distributions and their means and variances. We also present approximate solutions for the whole model, complementing with exact stochastic simulations. We show that bursts generally increase the noise in organelle numbers, producing distinct signatures in noise profiles depending on different mechanisms of organelle biogenesis. We also find different shapes of organelle number distributions, including bimodal distributions in some parameter regimes. Notably, bursty synthesis broadens the parameter regime of observing bimodality compared to the `non-bursty' case. Together, our framework utilizes number fluctuations to elucidate the role of bursty synthesis in producing organelle number heterogeneity in cells.
 
\begin{description}
\item[Keywords] Organelle biogenesis, Stochastic process, Organelle number distribution

\end{description}
\end{abstract}

\maketitle


\section{\label{sec:level1}Introduction}

Eukaryotic cells possess several complex intracellular compartments called organelles, which serve fundamental functions such as cellular homeostasis, metabolism, growth, and division \cite{marshall2020scaling,marshall2016cell,julian_organelle_lysosome,burbridge2022organelle}. Though intricate molecular pathways maintain the number of specific organelles in a cell, mechanistically, four basic processes control the organelle abundance (see Fig. 1) \cite{Mukherji_elife,choubey_organelle,craven_elife,marshall2016cell}. Most organelles can form \textit{de novo} from a preexisting membrane source and decay through maturation or autophagy. Moreover, multiple organelles can fuse together and may also undergo fission. Some of these processes, however, predominate over others for a specific type of organelles. For example, the abundance of Golgi apparatus \cite{rossanese_golgi,bevis_golgi,losev_golgi,matsuura_golgi} and lipid droplets \cite{pol_lipidDroplet,walther2017lipid} are maintained by \textit{de novo} synthesis and decay. On the other hand, mitochondria \cite{diaz_mitochondrial,cerveny_mitochondria_regulation,mitochondrial_mechanisms} and vacuoles \cite{wickner_vacuole,Mayer_vacuole,wickner_vacuole_review} are mainly affected by fission and fusion; whereas peroxisome's number is regulated via \textit{de novo} synthesis, fission, and decay \cite{motley_peroxisome,vandarZand_peroxisome,hoepfner_peroxisome}.

\par
Notably, the above processes of organelle biogenesis occur stochastically, producing cell-to-cell variability in organelle abundance when the number of organelles is low inside a cell. One of the fundamental questions in cell biology is how the cell regulates the variability in organelle numbers, and it has emerged as an important area of research recently \cite{Mukherji_elife,craven_elife,choubey_organelle,chang_marshall_heterogenity}. It is interesting to note that large fluctuations in mRNA and protein copy numbers also arise during gene expression, which has long been studied \cite{sanchez_gene_expr_rev,iyer_stochasticity,das2017effect, friedmanPRL}.  In these cases, the framework of stochastic processes has been helpful in drawing theoretical predictions for experimental verification. 

\par
Depending on various biological contexts, additional sources of variability can be underlying the organelle biogenesis processes. For example, it has been experimentally observed that subcellular structures like cilia and flagella grow from bursts of random sizes of building blocks  \cite{ludington_avalanche,bauerIScience}. A recent study reports that bursts can happen during the growth of many organelles such as Golgi bodies, peroxisomes, lipid droplets, and mitochondria \cite{Mukherji_PRL}. However, bursts occur not only in organelle size, but there could be burst in organelle number during the \textit{de novo} synthesis process. For instance, an experiment suggested that peroxisome synthesis in mouse liver cells can happen in bursts during embryonic development \cite{ontogeny_peroxisome_mouse-liver}. Some studies also reported that \textit{de novo} synthesis of centrioles occurs in bursts which can lead to the large variability of their numbers ranging from $1$ to $14$ centrioles per cell \cite{marshall2016cell,marshall_centriole, Terra_centriole,khodjakov_centriole}. 

\par
In the above context, `burst' generally means a rapid increase of a stochastic variable (either organelle size or number or both) during short time intervals. Such bursty phenomena have also been reported in mRNA and protein production during gene expression \cite{cai_stochastic,sanchez_gene_expression_burst,friedmanPRL,Kulkarni_burstyGeneExpr,swain_analytical,kumar2015transcriptional, jia2017simplification_Markov-chain, mackey2013dynamic_SIAM}. The burst size distribution (i.e., the distribution of organelle numbers produced per burst) can only be determined in experiments and is not always known. We can, nevertheless, presume a reasonable shape of the burst size distribution. A heuristic guess is an exponential or geometric distribution, implying that a large burst size is less likely than a low one. In fact,  experimental studies on noisy gene expression have found the burst size distribution of gene products to be exponential \cite{cai_stochastic}. However, in the case of organelles, the nature of the burst size distribution is not known. Recently, Amiri \textit{et. al.} \cite{Mukherji_PRL} have hypothesized that organelles can grow in exponentially distributed random sizes. Here, we want to investigate an alternative hypothesis: the implication of geometrically distributed bursts in number during organelle synthesis.

\par
Inspired by the observed bursty dynamics in organelles \cite{ontogeny_peroxisome_mouse-liver,ludington_avalanche,bauerIScience, Mukherji_PRL,marshall2016cell}, we thus asked: How does bursty {\it de novo} synthesis generally affect the organelle number distribution?
  We considered the organelle biogenesis model of Mukherji \textit{et. al.} \cite{Mukherji_elife} and modified it by making the \textit{de novo} synthesis bursty. We aimed to solve the model analytically in the steady state and calculate the probability distribution of organelle numbers and their mean and variance. We show that our model can be solved exactly in different limiting cases. Moreover, for the full model, we provide an approximate expression of probability distribution and relevant moments of the organelle number. We also compared our analytical results with stochastic simulations to validate our theoretical predictions.
  
\par
Our analysis suggests that bursty synthesis generally enhances noise in organelle abundance. In the absence of bursts, it was known that the interplay between \textit{de novo} synthesis and fission produce bimodal distributions of organelle numbers \cite{choubey_organelle}. However, bursty synthesis broadens the parameter regime of observing bimodality. As a result, in some parameter regimes, the unimodal distribution observed in the `no-burst' case becomes bimodal due to the influence of bursty synthesis. Moreover, different processes of organelle biogenesis lead to distinct signatures in the variability of organelle number (captured by the Fano factor). Together, bursts enhance cell-to-cell diversity in organelle abundance.

\begin{figure}[hbt!]
\begin{center}
{\includegraphics[height=6cm]{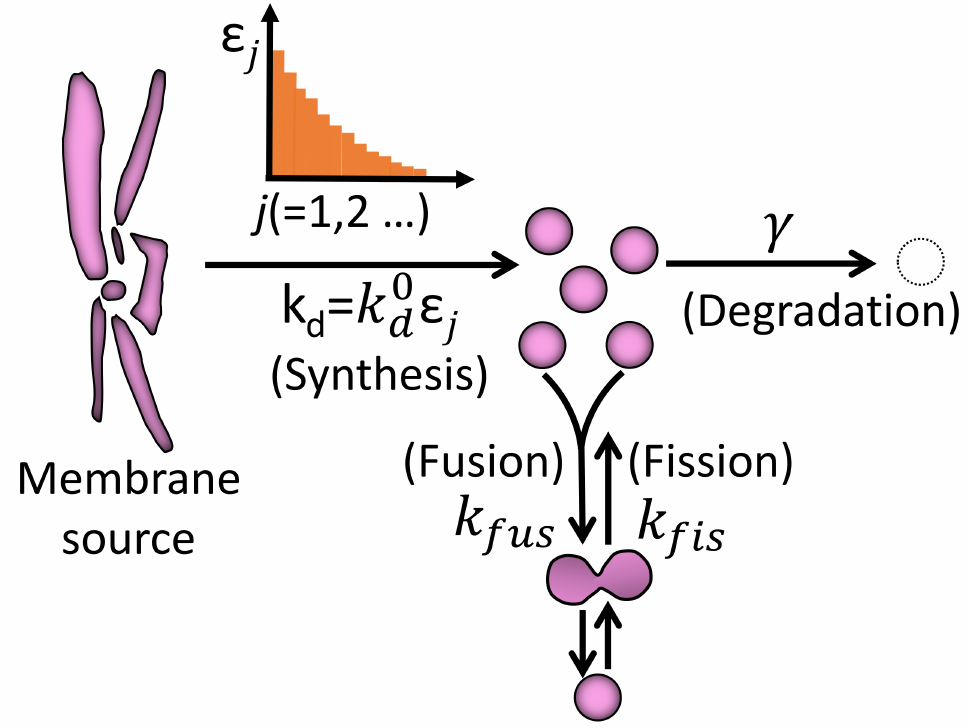}
\caption{Model of organelle biogenesis with bursty synthesis. Organelle biogenesis involves four processes, namely, \textit{de novo} synthesis, fission, fusion, and decay, which occur with rates $k_d$, $k_{fis}$, $k_{fus}$, and $\gamma$ respectively. The number of organelles produced per burst, $j$, obeys a geometric distribution ($\epsilon_j$), and $k_d$ is modified accordingly (see Model).
}}\label{fig1}
\end{center}
\end{figure}

\section{\label{sec:level1}Model}

We adapted a published model of organelle biogenesis \cite{Mukherji_elife,craven_elife,choubey_organelle,Choubey_temporal} by including a `bursty' \textit{de novo} synthesis (see Fig. 1).
 \textit{De novo} synthesis means production from scratch. Through bursty \textit{de novo} synthesis, more than one organelle can be randomly produced at a time. Here, we considered the number of organelles synthesized per burst to be geometrically distributed, motivated by earlier experiments \cite{cai_stochastic,Mukherji_PRL}. Such assumption of geometric (or exponential) burst size distribution was also considered in theoretical studies of bursty gene expression \cite{paulsson2000,friedmanPRL}. In our model, if we define q $(0< q <1)$  as the \textit{a priori} probability of producing one organelle, then $\epsilon_{j}=(1-q){q}^j$ can be taken as the likelihood of a burst of $j$ organelles. Considering $k^0_d$ as the rate of initiating production, the \textit{de novo} synthesis occurs at a rate $k_d=k^0_d\epsilon_{j}=k^0_d(1-q){q}^j$. Here, $k^0_d$ can be thought of as the bare rate of \textit{de novo} synthesis in the absence of any burst (i.e, when $\epsilon_1=1 \hspace {1mm}\text{and} \hspace{1mm} \epsilon_{j>1}=0 $). Moreover, fission and decay take place with rates  $k_{fis}$ and $\gamma$, respectively. These processes are first order since they depend on instantaneous organelle abundance. On the other hand, fusion is a second-order process happening at a rate $k_{fus}$.

\section{\label{sec:level1}Method}

  We analyzed the above model using the framework of stochastic process \cite{paulsson2000}. We define $P(n,t)$ as the probability of having $n$ organelles at the time $t$. Then, the Master equation describing the time evolution of $P(n,t)$ becomes

 \begin{align}
   &\frac{dP(n,t)}{dt} = k^0_d\sum_{n^{'}=0}^{n-1} (1-q){q}^{n-n^{'}} P(n^{'},t)+ \gamma(n+1)P(n+1,t)  \nonumber \\ &+ k_{fis}(n-1)P(n-1,t) + k_{fus}n(n+1)P(n+1,t)  \nonumber \\ &- \Biggl [k^0_d\sum_{n^{'}=n+1}^{\infty} (1-q){q}^{n^{'}-n} + \gamma n + k_{fis}n \nonumber \\ & + k_{fus}n(n-1)\Biggr]P(n,t).
\end{align}
 
 Here, $q$, $k^0_d$, $\gamma$, $k_{fis}$,  and $k_{fus}$ are positive real constants as defined in the model (and, $0< q <1$). We solved the Master equation (Eq. (1)) using the generating function method to obtain the steady-state number distribution of organelles, $P(n)$ $\left(\textit{i.e.,}~\text{when}~{dP(n,t)}/{dt}=0\right)$.
 
 \par
 Complementary to our analytical approach, we also used the Gillespie algorithm \cite{gillespie1977exact} to simulate the processes of organelle biogenesis. All the kinetic processes can be summarised by the following set of reactions (see Fig. 1):
 \begin{itemize}
 \item [(i)] n$\xrightarrow{k^0_d\epsilon_{j}}$ n+j  (j=1,2,$\cdots$) (Bursty \textit{de novo} synthesis)\\
 \item [(ii)] n$\xrightarrow{k_{fis}n}$ n+1   (Fission) \\
 \item [(iii)] n$\xrightarrow{k_{fus}n(n-1)}$ n-1    (Fusion) \\
 \item [(iv)] n$\xrightarrow{\gamma n}$ n-1    (Decay)
 \end{itemize}
From stochastic simulations, data were analyzed at large times, so that mean and variance become independent of time, ensuring a steady state.

\section{\label{sec:level1}Results}

Before solving the full model, we first focused on the possible limiting cases since, for specific organelles, some processes dominate over others. Different possibilities exist for combining the four basic processes (\textit{de novo} synthesis, fission, fusion, and decay) through which cells can regulate organelle abundance. We obtained exact analytical results in some limits involving the bursty synthesis, namely (i) Bursty synthesis-decay, (ii) Bursty synthesis-fusion, (iii) Bursty synthesis-fission-decay, and (iv) Bursty synthesis-fusion-decay. It is worthwhile to mention that combining fission-fusion (without involving {\it de novo} synthesis) also has a steady state solution shown earlier \cite{craven_elife,choubey_organelle}. We first discuss the limiting cases below.

\subsection{\label{sec:level2}Bursty synthesis-decay}

The limiting case of bursty synthesis and decay can be relevant for organelles like Golgi bodies \cite{rossanese_golgi,bevis_golgi,losev_golgi,matsuura_golgi} and lipid droplets \cite{pol_lipidDroplet,walther2017lipid}, which are maintained primarily by \textit{de novo} synthesis and decay processes. Also note that this limiting submodel boils down to a previously analyzed model of bursty gene expression \cite{cai_stochastic, paulsson2000, jia2018relaxation}. 

In the absence of burst, the synthesis-decay submodel is a standard `birth-death' process resulting in a Poisson distribution of organelle numbers \cite{craven_elife,choubey_organelle}. Bursty synthesis modifies the distribution to the negative binomial (see SI, Section S1). The probability of having $n$ number of organelles is  
\begin{align}
P(n)= \frac{\Gamma(n+r)}{n! \Gamma(r)}(1-q)^r q^n \hspace{2mm} \text{where,}\hspace{2mm} r=\frac{k^0_d}{\gamma},
\end{align}
and the corresponding mean and variance are
\begin{align}
   &\langle n\rangle=\frac{rq}{1-q} \\& \sigma^2=\frac{rq}{(1-q)^2}.
\end{align}

Also note that $B = q/(1-q)$ is the average burst size, i.e., the mean of the geometric distribution ($\epsilon_{j}=(1-q){q}^j$). Therefore, $\langle n \rangle = rB$ and $\sigma^2 = rB(1+B)$. Here, $r = k^0_d / \gamma$ is frequently interpreted as the `burst frequency' during bursty gene expression \cite{jia2019single-cell}. 
Next, we choose the Fano factor, defined as the ratio of variance to the mean, as a measure of noise in organelle number to understand the deviation from the Poisson distribution (for which Fano factor $=1$). From Eq. (3) and (4), we see that the 
Fano factor, $ \sigma^2/\langle n\rangle =1/(1-q)=1+B$, is independent of the synthesis and decay rates. Since $q$ is always less than 1, Eq. (2) is a super-Poissonian distribution (for which fano factor $>1$). As shown in Fig. 2A, for a fixed mean, the noise increases with higher values of $q$ (i.e., when the likelihood of producing more than a single organelle per burst increases).

\subsection{\label{sec:level2}Bursty synthesis-fusion}
All the statistical quantities of the `bursty synthesis-fusion' submodel can be described by two parameters, namely $\xi=k^0_d/k_{fus}$ (the ratio of \textit{de novo} synthesis to fusion rate) and $q$. 
We calculated the distribution of organelle numbers given by (see SI, Section S2)
 \begin{align}
     &P(n)= \frac{1}{_2F_1\left(a,b;c;q\right)} \frac{(a)_{n-1}(b)_{n-1}}{(c)_{n-1}} \frac{q^{n-1}}{(n-1)!} \hspace{2mm} \text{and} \nonumber \\ &P(0)=0,
\end{align}
  where $(a)_n=a (a+1) (a+2)...(a+n-1) $ is the Pochhamer symbol, and  $a=\frac{1}{2}(1+\sqrt{1-4\xi})$, $b=\frac{1}{2}(1-\sqrt{1-4\xi})$ and $c=2$. Note, since there is no decay, the organelle number cannot go below one (hence $P(0)=0$). 
  
  We next calculated the mean and variance of the organelle numbers given by \\
  \begin{align}
   &\langle n\rangle = 1+\frac{qab \hspace{1mm} _2F_1\left(1+a,1+b;1+c;q\right)}{c\hspace{1mm} _2F_1\left(a,b;c;q\right)} ,\hspace{1mm}\text{and}\hspace{1mm} 
   \end{align}

  \begin{align} 
 \sigma^2 = & q a b \Biggl[ \hspace{1.5mm}_2F^R_1\left(a,b;c;q\right) \Biggl\{\frac{_2F^R_1\left(1 + a,1 + b;1 + c;q\right)}{_2F^R_1\left(a,b;c;q\right)^2} \nonumber \\ &  + \frac{q\left(1 + a\right) \left(1+b\right) _2F^R_1\left(2 + a,2 + b;2 + c;q\right)}{_2F^R_1\left(a,b;c;q\right)^2} \Biggr\}\Biggr] \nonumber \\ &  - \left[\frac{q a b\hspace{1mm} _2F^R_1\left(1+a,1+b;1+c;q\right)}{_2F^R_1\left(a,b;c;q\right)}\right]^2,
\end{align}
 where, $_2F_1\left(a,b;c;q\right)$ is the hypergeometric function, and $_2F^R_1\left(a,b;c;q\right)={_2F_1\left(a,b;c;q\right)}/{\Gamma(c)}$ is the regularized hypergeometric function.
  
  We further note that the `synthesis-decay' submodel differs from the `synthesis-fusion' submodel in how the organelle number reduces over time. In these models, organelle numbers decrease via decay or fusion, respectively. Since the decay process depends linearly on the instantaneous organelle number and fusion depends quadratically, it is expected that fusion may suppress noise in the later model. In the absence of burst, previous studies \cite{Mukherji_elife,choubey_organelle} showed that the `synthesis -fusion' model leads to sub-Poisson noise ( i.e., Fano factor $< 1$ always). However, the inclusion of burstiness can make the noise go above one, as shown in Fig. 2B (see the case of $q=0.7$). Therefore, bursts (with high $q$) can produce super-Poissonian behavior (Fano factor $>1$) in contrast to sub-Poissonian behavior in the absence of bursts.

  \begin{figure*}[ht!]
\begin{center}
{\includegraphics[height=10cm]{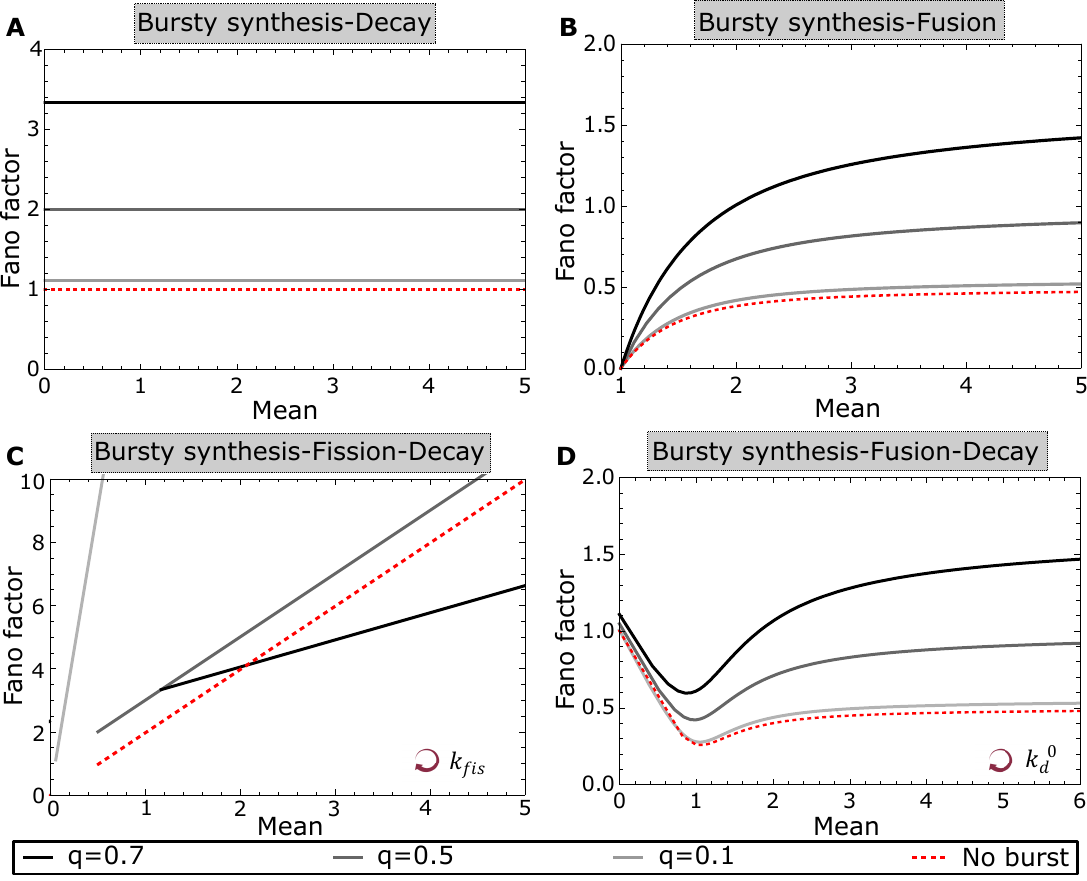}
\caption{Fano factor versus mean plots for different limiting cases of our model. (A) The Fano factor is independent of the mean and only depends on $q$ for the `bursty synthesis-decay' submodel. (B) For the `synthesis-fusion' submodel, the parametric curves for Fano factors were obtained by varying the ratio of \textit{de novo} synthesis rate ($k^0_d$) to fusion rate ($k_{fus}$). (C) The fission rate ($k_{fis}$) was varied for the `synthesis-fission-decay' submodel to generate the noise profiles while keeping other rates fixed (here $k^0_d=0.5$ and $\gamma=1$). (D) We tuned $k^0_d$ for the `synthesis-fusion-decay' submodel and kept other rates fixed ($k_{fus}=50$ and $\gamma=1$). All rate parameters are in arbitrary units of time$^{-1}$. In all panels, different shades of grey represent different $q$ values, and red dashed curves represent the `no-burst' cases (in `no burst' cases, the synthesis happens at a constant rate $k^0_d$ without involving any burst size distribution \cite{choubey_organelle}). All curves are plotted from exact analytical expressions.
}}\label{fig2}
\end{center}
\end{figure*}

  

\subsection{\label{sec:level2}Bursty synthesis-fission-decay}

The limiting submodel of `bursty synthesis-fission-decay' can be relevant for peroxisome abundance, which is known to be regulated mainly by \textit{de novo} synthesis, fission, and decay \cite{motley_peroxisome,vandarZand_peroxisome,hoepfner_peroxisome}. Note that Bursty synthesis and fission increase the number of organelles while the number reduces via decay. Since fission increases organelle number exponentially, a steady state can only be reached when the fission rate is less than the decay rate ($k_{fis}<\gamma$) --- this was also shown before \cite{choubey_organelle} in the `no-burst' case. 
 
\par
In our case, the `bursty synthesis-fission-decay' submodel can be fully characterized by three parameters:  $\alpha=k^0_d/k_{fis}$, $\eta=k_{fis}/\gamma$, and $q$. We obtained the steady-state probability distribution of organelle numbers as  

\begin{align}
P(n)= &\left[\frac{1-\eta}{1-q}\right]^m \frac{\Gamma(n+m)}{\Gamma(m)n!}\eta^{n}. \nonumber\\ &
\hspace{1mm}{_2F_1\left(-n,-m;1-n-m;\frac{q}{\eta}\right)},
\end{align}
where $m=\left(\frac{\alpha \eta q}{\eta - q}\right)$ (see SI, Section S3). In contrast, the organelle number distribution for this limiting model is negative binomial in the `no-burst' scenario \cite{craven_elife,choubey_organelle}. We next calculated the mean and variance as

\begin{align}
   & \langle n\rangle=\frac{q \alpha \eta}{(q-1)(\eta-1)}\\&
   \sigma^2=\frac{q \alpha \eta (1-q\eta)}{(q-1)^2(\eta-1)^2}.
\end{align}

Note that since $q$ is always less than one, for a positive mean,  $\eta (=k_{fis}/\gamma)$ has to be always less than one, i.e., $k_{fis}<\gamma$. This is, in fact, the condition for achieving a steady state.

We further explored the behavior of the Fano factor as a function of the mean by varying only one of the rate constants (among $k_{fis}$, $\gamma$, and $k^0_d$), keeping the other two fixed. Here, we varied $k_{fis}$ since the fission rate for peroxisomes in yeast can be tuned by genetic knockdowns \cite{peroxisome_gene_knockdown}. Finally, using Eq. (9) and (10), the Fano factor can be expressed as
 \begin{align}
 \frac{\sigma^2}{\langle n\rangle} = \left(\frac{\gamma}{k^0_d}\right)  \frac{\left(1-q\right)}{q} \langle n\rangle + \frac{q}{(1-q)} \nonumber. 
 \end{align}
 Thus, the Fano factor is a linearly increasing function of the mean (see Fig. 2C). As $B = q/(1-q)$ is the average burst size, the above equation for the Fano factor can be written as 
 \begin{align}
 \frac{\sigma^2}{\langle n\rangle} = \left(\frac{\gamma}{k^0_d}\right) (1/B) \langle n\rangle + B \nonumber. 
 \end{align}
 Here, the average burst size, $B$, determines slopes and intercepts (when $\gamma$ and $k^0_d$ are fixed) of the straight lines representing the Fano factor versus mean plots (see Fig. 2C).
 In particular, the slopes are inversely proportional to B.


\subsection{\label{sec:level2}Bursty synthesis-fusion-decay}

The `Bursty synthesis-fusion-decay' limit of our model can be relevant for vacuoles \cite{wickner_vacuole,Mayer_vacuole,wickner_vacuole_review} for which fusion plays a determining role. This submodel is delineated by three parameters, $\delta=\gamma/k_{fus}, \xi=k^0_d/k_{fus}$, and $q$. We derived the  organelle number distribution as (see SI, Section S4)
 \begin{align}
     P(n)= \frac{1}{_2F_1\left(a',b';c';q\right)} \frac{(a')_{n}(b')_{n}}{(c')_{n}} \frac{q^{n}}{n!},
 \end{align}
 and corresponding mean and variance as, \\
 \begin{align}
     \langle n\rangle = \frac{qa'b' \hspace{1mm} _2F_1\left(1+a',1+b';1+c';q\right)}{c'\hspace{1mm} _2F_1\left(a',b';c';q\right)} 
\end{align}

\begin{align}
\sigma^2 = & q a' b' \Biggl[ \hspace{0.5mm}_2F^R_1\left(a',b';c';q\right) \Biggl\{\frac{_2F^R_1\left(1 + a',1 + b';1 + c';q\right)}{_2F^R_1\left(a',b';c';q\right)^2} \nonumber \\ &  + \frac{q\left(1 + a'\right) \left(1+b'\right) _2F^R_1\left(2 + a',2 + b';2 + c';q\right)}{_2F^R_1\left(a',b';c';q\right)^2} \Biggr\}\Biggr] \nonumber \\ &  - \left[\frac{q a' b'\hspace{1mm} _2F^R_1\left(1+a',1+b';1+c';q\right)}{_2F^R_1\left(a',b';c';q\right)}\right]^2,
\end{align}

 where, 
 \begin{align*}
 & a'=\frac{1}{2}\left[\delta - 1 + \sqrt{(\delta - 1)^2-4\xi}\right] \nonumber \\ 
 & b'=\frac{1}{2}\left[\delta - 1 - \sqrt{(\delta - 1)^2-4\xi}\right] \text{and} \hspace{1.5mm}c'=\delta.
 \end{align*}

 We next explored the behavior of the Fano factor with mean by varying $k^0_d$ (keeping $k_{fus}$ and $\gamma$ fixed). As shown in Fig. 2D, we observed non-monotonic behaviors in the Fano factors, showing a prominent dip near the mean $\sim 1$. This non-monotonic behavior stems from the fact that at least two organelles are required for fusion to take place, and we used a much higher fusion rate than the decay rate ($k_{fus}=50$ and $\gamma=1$). Notably, the instantaneous rate of fusion is $k_{fus} n (n-1)$, where $n$ is the instantaneous organelle number. This term contributes only when $n \geq 2$. Thus, the cross-over from a `No-fusion'  regime (for mean $<$ 2) to a fusion-dominated regime (for mean $>$ 2) produces the observed non-monotonic behavior. Note that the fusion starts dominating as soon as $n$ exceeds $2$ since $k_{fus} \gg \gamma$ in our case. In fact, after the minima in the Fano factor near mean $\sim 1$, the Fano factor reaches a super-Poissonian limit for high $q$, and the later parts of the curves in Fig. 2D (above mean $\sim 2$) are almost the same as the curves shown in Fig. 2B, suggesting the dominance of `synthesis-fusion' processes over decay. On the other hand, if $k_{fus} \ll \gamma$, the dominance of fusion would build up at very high values of $n$, and the observed dip in Fano factors vanish with decreasing $k_{fus}$, compared to $\gamma$ (see Fig. S1). Eventually, when $k_{fus} \ll \gamma$, the initial part of the Fano factor curves (near zero mean) become similar to the `synthesis-decay' submodel where the presence of fusion cannot be perceived (compare Fig. 2A with Fig. S1C).
 
  Interestingly, for the `bursty synthesis-fusion-decay' submodel, following Jia \textit{et al.} \cite{jia2017stochastic} we can decompose the Fano factor as 
 \begin{align}
 FF = FF_{\text{syn-decay}}-\Delta FF, \nonumber
 \end{align}
 where $FF_{\text{syn-decay}}=1+B$ is the Fano factor of the `bursty synthesis-decay' submodel, and $\Delta FF =({Cov(n,f_n)-(1+B)\langle f_n \rangle})/({k^0_d B-\langle f_n \rangle})$ is the amount of noise reduced by the fusion process (see SI subsection S4.1 ). Here, $B=q/(1-q)$ is the average burst size, $f_n = k_{fus}n(n-1)$ is the fusion propensity, and $ Cov(n,f_n)= \langle nf_n \rangle - \langle n \rangle \langle f_n \rangle $ is the covariance between $n$ and $f_n$. We verified $\Delta FF >0$ (see Fig. S2). Thus, the inclusion of fusion suppresses the noise from the `bursty synthesis-decay' submodel, and hence, fusion acts like negative feedback.

\begin{figure*}
\begin{center}
{\includegraphics[width=0.82\textwidth,height=8.8cm]{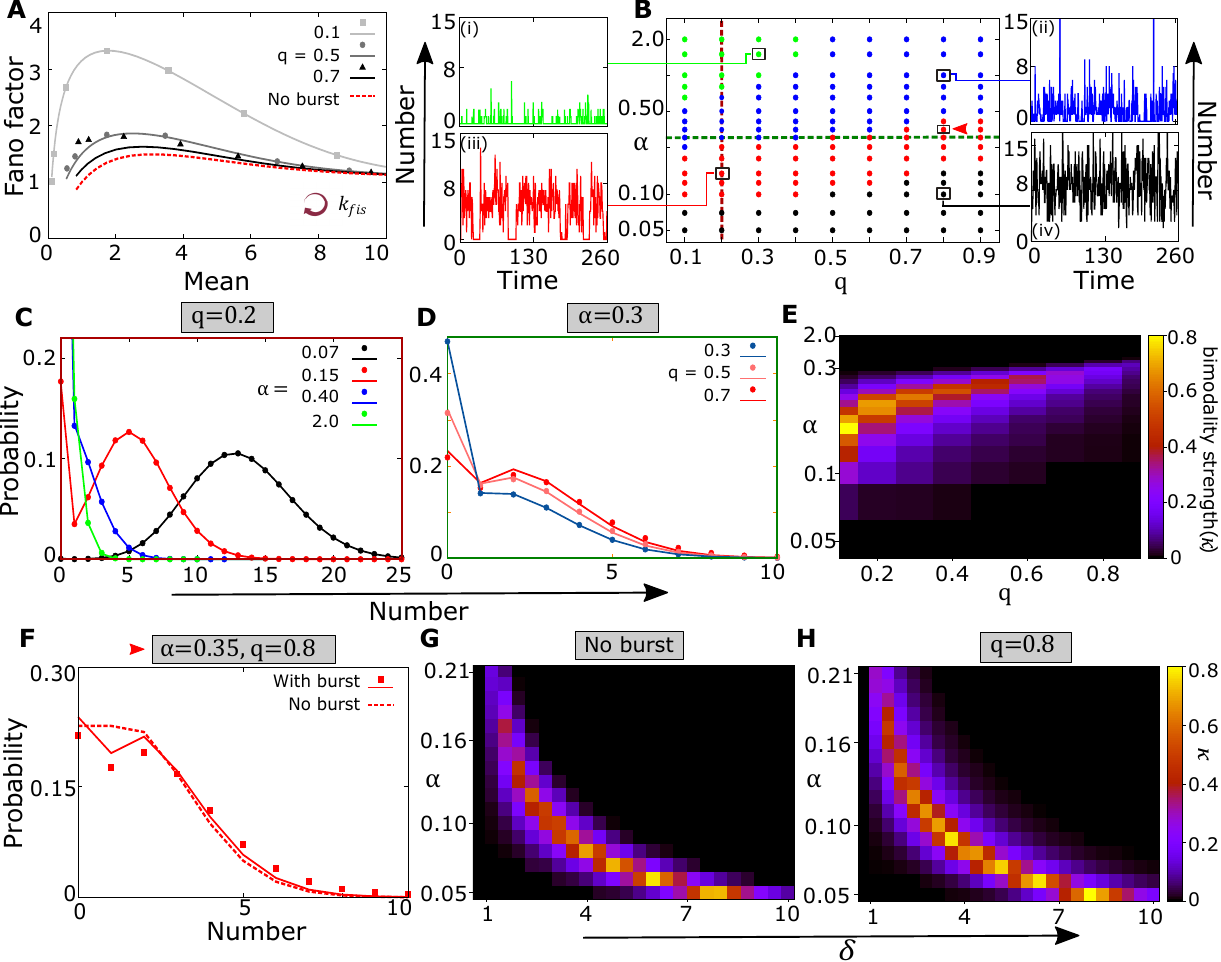}
\caption{Effect of bursty synthesis in the full model. (A) Fano factor versus mean curves are shown by varying $k_{fis}$ for different values of $q$. (B) Distinct dynamical regimes showing different types of organelle number distributions. The green, blue, red, and black dots represent exponential, long-tailed, bimodal, and unimodal distributions, respectively,  in the parameter space of $\alpha=k^0_d/k_{fis}$ and $q$. Corresponding time traces of organelle numbers (produced by Gillespie simulation) are also shown in panels (i-iv). The brown dashed vertical line and the green dashed horizontal line in the main panel of B correspond to panels C and D, where representative distributions are shown for chosen parameter values. (C) At a fixed $q=0.2$ (corresponding to the brown dashed vertical line in B), with increasing $\alpha$, organelle number distributions change from unimodal to exponential via bimodal and long-tailed shapes. (D) Conversely, at a fixed $\alpha=0.3$ (corresponding to the green dashed horizontal line in B), distributions become long-tailed to bimodal with increasing $q$. (E) The heatmap in the $q - \alpha$ parameter space shows the strength of bimodality (corresponding to Panel B) calculated by the parameter $\kappa$ (defined in Eq.  20).(F) A comparison of bimodal and unimodal distributions in the bursty and non-bursty cases, respectively. The red arrowhead in panel B ($\alpha=0.35$ and $q=0.8$) indicates the point where the distribution was calculated for the bursty case. (G, H) Comparative heatmaps of the bimodality strengths in the parameter space of $\alpha$ and $\delta (=\gamma/k_{fus})$, showing that the bimodality region broadens with bursts compared to the `no-burst' case. For panels A-F, we kept  $k^0_d=k_{fus}=\gamma=1$ and for G-H, $k^0_d=k_{fus}=1$. In panels A, C, D, and F, points represent simulation data, while solid (or dashed) curves are from analytical expressions. }}\label{fig3}
\end{center}
\end{figure*}


\subsection{\label{sec:level2}Full model: Bursty synthesis-fission-fusion-decay}

After investigating different submodels, we finally proceed to analyze the general model with all four processes and explore the impact of bursty synthesis on organelle number distribution. For the full model, we started from the general Master equation (Eq. (1)) and derived the equation of the probability generating function (defined as $G(z)=\sum_{n=0}^{\infty} z^{n} P(n)$):
  \begin{align}
 z\derivative[2]{G(z)}{z} + (\delta - \phi z) \derivative{G(z)}{z} - \alpha \phi \frac{q}{1-qz}G(z)=0 ,
 \end{align}
 where $\alpha=k^0_d/k_{fis},  \delta=\gamma/k_{fus}$, and $\phi=k_{fis}/k_{fus}$
 are positive real constants (see SI, Section S5). The singularity analysis of Eq. (14) shows that it has two regular singular points and one irregular singular point at infinity, and thus
this differential equation is not of Riemann type. Hence, we could not convert the equation to the Gauss hypergeometric form, enabling the solution in terms of hypergeometric functions. However, we circumvented this issue by approximating the actual generating function equation to an extended confluent hypergeometric equation \cite{campos2001some} (see SI, Section S6 for details). We approximated Eq. (14) as follows.   When  $|qz|<1$ (also note, $q<1$), Eq. (14) can be expressed as  
 \begin{align}
 z\derivative[2]{G(z)}{z} + (\delta - \phi z) \derivative{G(z)}{z} - \alpha \phi q\sum_{k=0}^{\infty}(qz)^kG(z)=0
 \end{align}
Neglecting higher order terms in the summation over $k$ (considering
$k$ up to 1), Eq. (15) can be reduced to 
 \begin{align}
 z\derivative[2]{G(z)}{z} + (\delta - \phi z) \derivative{G(z)}{z} - \alpha \phi q(1+qz)G(z) \approx 0,
 \end{align}
 which can be solved in terms of hypergeometric functions. Finally, we obtained the approximate probability distribution as 
\begin{align}
 P(n) \approx \frac{({\frac{\mu}{2}})^n}{\exp{\frac{\mu }{2}}{_1F_1\left(\kappa,d;\zeta \right)} }\frac{_2F_1\left(-n,\kappa;d;-\frac{2\zeta}{\mu}\right)}{n!}.
 \end{align}
The corresponding approximate mean and variance are
 \begin{align}
     \langle n\rangle  \approx \mu + \frac{\kappa \zeta\hspace{1mm} _1F^R_1\left(1+\kappa,1+d;\zeta\right)}{ _1F^R_1\left(\kappa,d;\zeta\right)} 
 \end {align}

 \begin{align}
 \sigma^2 \approx &  \mu + \kappa \zeta \Biggl[ \hspace{1.5mm}_1F^R_1\left(\kappa,d;\zeta\right) \Biggl\{\frac{_1F^R_1\left(1 + \kappa,1 + d;\zeta\right)}{_1F^R_1\left(\kappa,d;\zeta\right)^2} \nonumber \\ &  + \frac{\zeta\left(1 + \kappa\right)  _1F^R_1\left(2 + \kappa,2 + d;\zeta\right)}{_1F^R_1\left(\kappa,d;\zeta\right)^2} \Biggr\}\Biggr] \nonumber \\ &  - \left[\frac{\kappa \zeta \hspace{1mm} _1F^R_1\left(1+\kappa,1+d;\zeta\right)}{_1F^R_1\left(\kappa,d;\zeta\right)}\right]^2
 \end{align}
where $\Tilde{\alpha}=\alpha\phi q$, $A_1=\alpha\phi q^2$, $\zeta=\sqrt{\phi^2+4A_1}$, $d=\delta$, $\mu=\left(\phi-\zeta\right)$, and $\kappa=\left({\Tilde{\alpha}-\frac{d}{2}\mu} \right) / {\zeta}$ (see detailed calculations in SI, Section S7). Also, $_1F_1\left(\kappa,d;\zeta\right)$ is the confluent hypergeometric function, and $_1F^R_1\left(\kappa,d;\zeta\right)={_1F_1\left(\kappa,d;\zeta\right)}/{\Gamma(d)}$ is the regularized confluent hypergeometric function.

As before, we explored the behavior of the Fano factor by varying the fission rate $k_{fis}$, keeping other parameters fixed (here $k^0_d=k_{fus}=\gamma=1$). The Fano factor exhibited non-monotonic behavior with mean, showing a peak near mean $\approx 2$ (see Fig. 3A). The origin of this non-monotonic behavior may be understood by noting that fusion can not occur when the organelle number is less than 2. Thus, bursty synthesis, fission, and decay processes dominate at the low organelle numbers. Conversely, `fission-fusion-decay' processes dominate at high organelle numbers since the fission rate becomes much higher than the synthesis rate. Moreover, since \textit{de novo} synthesis is a zeroth order process and fission is a first-order process (see reactions (i) and (ii) in Model), the propensity of fission becomes much higher at high organelle numbers (i.e., $k_{fis} n \gg k^0_d \epsilon_j$). Together, the peak in the Fano factor signifies a cross-over from the `synthesis-fission-decay' dominated regime to the `fission-fusion-decay' dominated regime (also see Fig. S3 in Section S8). The peak suggests that the organelle number distribution near the cross-over regime may become much broader and pass through interesting shapes. 

In fact, a previous study  \cite{choubey_organelle} reported that the interplay of synthesis and fission could produce bimodal organelle number distribution when the fission rate is higher than other parameters (i.e., $k_{fis}>\{k^0_d, \gamma, k_{fus}\}$). Note that fission can occur when the cell has at least a single organelle. However, if the organelle number goes to zero, it can again become nonzero only via {\it de novo} synthesis. Thus,
if the organelle number becomes zero stochastically, the cell stays long in this zero-organelle state when $k^0_d$ is much smaller than $k_{fis}$. Alternatively, the cell produces a finite organelle number once fission starts dominating for nonzero organelle numbers. This leads to two modes in the organelle number distribution, one at zero and the other at a nonzero number.

We next inspected how the bursty synthesis can affect the emergence of bimodal number distributions. We found different shapes of the number distributions, namely exponential, long-tailed, bimodal, and unimodal, in the parameter space of $\alpha = k^0_d/k_{fis}$ and $q$ (see Fig. 3B and 3C). When $\alpha$ is high, and $q$ is low, the organelle number stays near zero (see Fig. 3B(i)) since the fission rate is lower than the {\it de novo} synthesis rate, which itself is comparable to decay and fusion rates (i.e., $k_{fis} < \{k^0_d, \gamma, k_{fus}\}$ and $k^0_d \sim \gamma \sim k_{fus}$). Consequently, the number distributions become exponential (green dots in Fig. 3B and the green curve in Fig. 3C). In contrast, when the fission rate is higher than other rates ($k_{fis}>\{k^0_d, \gamma, k_{fus}\}$ and  $k_{fus} \lesssim \gamma$), the organelle numbers toggle stochastically between zero and nonzero values over time, as explained before (see time traces in Fig. 3B(iii)). Correspondingly, the distributions become bimodal (red dots in Fig. 3B and the red curve in Fig. 3C).  Moreover, distributions exhibit a transitory long-tailed shape between the `exponential' and `bimodal' regimes (blue dots in Fig. 3B). In this case, when $q$ is high and $k_{fis} \lesssim \{k^0_d, \gamma, k_{fus}\}$, the organelle number occasionally shoots up much higher than zero due to bursty synthesis, giving more weightage to the higher organelle number (see time traces in Fig. 3B(ii) and the corresponding blue curve in Fig. 3C). Finally,  when $\alpha$ is very low (i.e., $k_{fis} \gg \{k^0_d, \gamma, k_{fus}\}$), the organelle numbers are dominated by `fission-fusion-decay' processes, which eventually leads to a high organelle number on average and unimodal distributions (Fig. 3B(iv) and the corresponding black curve in Fig. 3C; also see Fig. S3 in Section S8).

Thus, for a fixed $q$, with decreasing $\alpha$ (or increasing $k_{fis}$ keeping other parameters fixed), the organelle number distribution can transit from exponential to bimodal via a long-tailed shape and ultimately reaches a unimodal shape (see Fig. 3C). Similarly, Fig. 3D shows that the organelle number distribution shifts from long-tailed to bimodal with increasing $q$ at a fixed $\alpha$. 
To quantify the shape of $P(n)$, following recent papers \cite{jia2020small,adhikary2023effects}, we further calculated the `bimodality strength' defined by 
\begin{align}
\kappa= (h_{low}-h_{valley})/h_{high}, 
\end{align}
where the heights of high and low peaks are denoted by $h_{high}$ and $h_{low}$ respectively, and $h_{valley}$ denotes the height of the valley between the peaks. It is evident that for bimodal distributions, $\kappa$ is a real number between 0 and 1, whereas, for unimodal distributions, it was set to 0. In Fig. 3E, we show the heatmap of $\kappa$ in the $q - \alpha$ parameter space. We found that the region of bimodality shrinks with increasing $q$ (Fig. 3E). 

Interestingly, we also found a parameter regime where the inclusion of bursty synthesis produces bimodal distribution, but the distribution was unimodal in the `no-burst'  case (Fig. 3F). Thus, to check whether bursty synthesis broadens the parameter regime of observing bimodality, we compared the heatmaps of bimodality strengths ($\kappa$) with and without bursts for $q=0.8$ (see Fig. 3G and 3H). For a fixed burst size, bursty synthesis indeed enhances the bimodal region compared to the `non-bursty' case in the parameter space of $\alpha$ and $\delta (=\gamma/k_{fus})$.

Note that our analytical expression of distribution (Eq. 17), although approximate, matches well with the simulation results despite showing slight deviations for high values of $q$ (Fig. 3C-D, and 3F). To estimate the deviation, we also calculated the relative errors in the approximate mean and variance in the $q$-$\alpha$ space (see Fig. S4). Note that the approximation for the generating function (Eq. 15 to 16) is poor when the product $\alpha \phi q$ is high. As expected, we observed significant errors in the high $q$ and high $\alpha$ regime. Otherwise, errors in the approximate mean and variance (compared to exact simulations) are much less than 15\%-20\% (Fig. S4). 
 

\section{\label{sec:level1}Discussion}
{\bf  Summary:} Bursts, i.e., rapid increase of some quantity in a short duration, are prevalent phenomena at the molecular and subcellular level. Examples of bursts include bursty protein expression \cite{cai_stochastic,sanchez_gene_expression_burst,friedmanPRL,Kulkarni_burstyGeneExpr,swain_analytical}, avalanche-like growth of flagella \cite{ludington_avalanche,bauerIScience}, and bursty growth of some organelles \cite{Mukherji_PRL}. Though further research is required to understand the implications of bursts, they may benefit organisms by producing phenotypic heterogeneity in rapidly changing environments \cite{stochastic_switching, plant_organellomic}. In this paper, we have investigated the effect of bursty organelle synthesis during organelle biogenesis. 

Burst provides a source of cell-to-cell variability in organelle number. Previous experiments suggest that some organelles, such as centrioles and peroxisomes, can be synthesized in bursts \cite{marshall2016cell,marshall_centriole, Terra_centriole,khodjakov_centriole,ontogeny_peroxisome_mouse-liver}. Here, we have proposed a model of organelle biogenesis with bursty {\it de novo} synthesis. We solved the model exactly in different limiting cases and presented analytical expressions of the steady-state probability distributions of organelle numbers, along with their means and variances. We considered the Fano factor (ratio of variance to mean) as a measure of the noise, which is generally enhanced by bursty synthesis. We further analyzed the full model (complemented by stochastic simulations) and found bimodal organelle number distributions in some parameter regimes. The bursty synthesis introduces a parameter $q$ (likelihood of producing one organelle per burst) that expands the parameter space of observing the bimodal distributions (compare the bursty and `non-bursty' cases in Fig. 3G and 3H). Consequently, in some parameter regimes, the organelle number distributions became bimodal with bursty synthesis but unimodal in the corresponding `non-bursty' case (Fig. 3F). Thus, bursty synthesis can enhance intracellular heterogeneity in organelle number.

We also note that the non-monotonic behavior of Fano factors against mean (as found in Fig. 3A) is similar to what is observed in models of genetic networks described by non-linear Master equations involving cross-talk between variables \cite{jiao2020cross-talk_BPJ, adhikary2023effects, das2017effect}. However, Fano factors may become monotonic if the underlying Master equation is linear in the stochastic variables of interest \cite{das2012giant}. 
Further, note that the Master equations in the absence of the fission process (described in Sections A, B, and D in Results) are special cases of the model described by Jia and Grima (see Eq. (3) in \cite{jia2020dynamical}), who proposed a reduced Master equation describing an auto-regulatory gene under fast switching approximation. Their model is essentially a
general `bursty birth-death' process, which can be mapped to our `bursty synthesis-fusion-decay' submodel by considering a birth rate $c_n = k^0_d$ and a death rate $d_n = k_{fus}n(n-1)+ \gamma n$. However, the full Master equation of organelle distribution with bursty synthesis (Eq. 1 in our paper) has not been rigorously studied before.

 An interesting limit of our full Master equation is how the bursty synthesis model (presented in this paper) can be reduced to the non-bursty model studied previously \cite{choubey_organelle}. Following the models on bursty protein synthesis \cite{jia2019single-cell,jia2020small,jia2023analytical_SIAM}, in the limit of $k_d^0 \longrightarrow \infty$ and $q \longrightarrow 0$,  while keeping $k_d^0 q=k_d$ as a constant, the equation for the probability generating function (Eq. 14), corresponding to our full Master equation,  turns out to be
\begin{align*}
 z\derivative[2]{G(z)}{z} + (\delta - \phi z) \derivative{G(z)}{z} - \frac{k_d}{k_{fus}} G(z)=0 .
\end{align*} 
The above equation is, in fact, the generating function's equation for the model without bursts, as derived previously \cite{choubey_organelle}.

{\bf Biophysical relevance: }
One can use our theoretical analysis of noise in organelle numbers to quantitatively unravel the effect of bursty mechanisms in the abundance of subcellular structures. Our analytical expressions for number distributions and Fano factors can help discern different mechanisms of organelle biogenesis, producing mechanistic insights. The results could be tested in microscopy experiments based on fluorescent proteins localized to specific organelles \cite{Mukherji_elife}. Our framework can also elucidate the role of bursts in physiological contexts. For instance, in yeast, peroxisomes proliferate under stress where the peroxisome number or size (or both) can rapidly increase from preexisting mature peroxisomes in a short duration of time \cite{yan_peroxisome_control, Mukherji_elife}. On the other hand, Kim \textit{et. al.} \cite{kim_mammelian-peroxisome_origin} observed that peroxisome synthesis occurs mainly via \textit{de novo} synthesis in mammalian cells . Thus, it could be an open question if a mammalian cell under stress can lead to bursty {\it de novo} synthesis of peroxisomes \cite{schrader_peroxisome_growth, kim_mammelian_perox-reviw}. Our predicted noise profiles can be a helpful tool to test such hypotheses.

Our observation of bimodal organelle number distributions further adds weight to the rising idea that phenotypic heterogeneity may arise through nongenetic means \cite{raser2005noise}, including the cell-to-cell variability of organelles \cite{chang_marshall_heterogenity}. In this context, our work suggests that bursty synthesis can be crucial in promoting such diversity.

{\bf Limitations and future direction:}
Even though we studied the role of bursty synthesis in organelle biogenesis theoretically, the burst size distribution, i.e., the organelle number produced per burst, is not experimentally measured. Inspired by bursty gene expression \cite{cai_stochastic, jia2023analytical_SIAM}, we chose a geometric burst size distribution since it makes intuitive sense that high burst sizes are less probable than small ones.

A possible origin of bursty organelle synthesis can be the multi-step nature of the synthesis that involves organelle maturation. As shown previously, protein bursts occur mainly due to the two-step protein production via mRNAs (i.e., transcription of mRNAs from a gene and translation of mRNAs into proteins), and since the intermediate species (mRNAs) are very short-lived \citep{swain_analytical}. Similarly, \textit{de novo} synthesis of organelles typically involves multiple steps, including the synthesis of membrane components, assembly of pre-mature structures and vesicular components, and finally maturation into functional structures \citep{glick2002can,van2012biochemically}. For example, pre-peroxisomal vesicles, which are derived from the endoplasmic reticulum (ER), combine by heterotypic fusions before assembling into matured peroxisomes \citep{van2012biochemically}. Also, vesicles budding from the ER sites fuse to form pre-Golgi elements, which further mature into Golgi bodies \citep{glick2002can}. If the lifetimes of such intermediate (pre-mature) vesicular components are much smaller than the matured organelles, then organelles can get synthesized in bursts (in a coarse-grained sense). However, exact kinetic rates of intermediate steps of organelle synthesis are not experimentally measured.

Moreover, all rate constants used in our stochastic simulations are arbitrary since they are not experimentally measured. Nevertheless, we show that our model (and its limiting sub-models) can be described by non-dimensional parameters, which are essentially ratios of various rate constants. Besides, our derived analytical formulas are valid for any positive real values of rate constants and can be tested in well-designed experiments. 

Here, we considered only those limiting cases that can be solved exactly at the steady state by incorporating bursty synthesis. We ignored the `bursty synthesis-fission-fusion' submodel despite having a steady state solution, as this cannot be solved exactly and since no known organelle corresponds to this sub-model (though this is a theoretical possibility). However, it is possible to obtain an approximate solution as was done for the full model. 

Furthermore, we here dealt with an analytically tractable model of organelle number regulation, ignoring detailed spatial structures of organelles. As shown recently, more complex theoretical models can describe specific spatiotemporal dynamics of subcellular structures, such as the inter-Golgi transport of proteins \cite{MRao-Golgi_PNAS, MRao-Golgi_ScReports}. A recent simulation also captured the evolution of both the size and number of organelles \cite{Mukherji_PRL}. Thus, in the future, it would be worthwhile to extend our model to include organelle sizes in an analytically tractable way. Also, the generating function equation for our full model (Eq. 14 or Eq. 15) represents an interesting example of a `non-Riemann-type' differential equation that exhibits distinct types of singularities and is challenging to reduce to any known form. The exact solution of this equation poses an intriguing mathematical problem.\\


\begin{acknowledgments}
We thank DBT (Government of India, Project No. BT/RLF/Re-entry/51/2018) and IISER-Kolkata for financial support. We thank CSIR. We also thank  Dibyendu Das (IIT Bombay), Shirshendu Chowdhury (IISER Kolkata), and Rajib Dutta (IISER Kolkata) for useful discussions.
\end{acknowledgments}

%


\nocite{*}



\clearpage
\widetext

\begin{center}
\textbf{\large Supplemental Materials: Effects of bursty synthesis in organelle biogenesis}
\end{center}

\setcounter{equation}{0}
\setcounter{figure}{0}
\setcounter{table}{0}
\setcounter{section}{0}
\makeatletter
\renewcommand\thefigure{S\arabic{figure}}
\renewcommand{\theequation}{S\arabic{equation}}
\renewcommand\thesection{S\arabic{section}}
 

\section{Bursty Synthesis-Decay}

 In this limit, the equation of generating function ($G(z)$) is  
 \begin{align}
  (1-z)\frac{dG}{dz}-rq\left[\frac{(1-q)z}{1-zq}-1\right]G=0 ,  
 \end{align}
  where $r=k^0_d/\gamma$ . Integrating over $z$ and using the normalization condition as the boundary condition  $G(z)=1$ at $z=1$, we get
  
  \begin{align}
      G(z)={(1-q)}^r{(1-qz)}^{-r}.
  \end{align}
  
  The $n^{\textit{th}}$ term in the expansion of $G(z)$ gives the probability of having $n$ number of organelles. Binomial expansion of Eq. (S2) provides Eq. (2),
\begin{align}
P(n)= \frac{\Gamma(n+r)}{n! \Gamma(r)}(1-q)^r q^n  \nonumber
\end{align}

 From the definition of generating function mean ($\langle n\rangle$) and variance ($\sigma^2$) can be written as 
 \begin{align}
 &\langle n\rangle=G'(1) \\& 
 \sigma^2=G''(1)+G'(1)-G'(1)^2,
 \end{align}
 where prime denotes derivative w.r.t. $z$. 
 
 In all cases, we calculated mean and variance using Eq. (S3) and (S4) and with the help of MATHEMATICA software.


\section{Bursty Synthesis-Fusion} 
 In this limit, the equation of generating function can be written as \\
 \begin{align}
 z\derivative[2]{G(z)}{z}-\left[\frac{\xi q}{1-zq}\right]G(z)=0,
 \end{align}
  where $ \xi=k^0_d/k_{fus}$ is a positive real constant. Here $z=0, 1/q$ are regular singular points and $z=\infty$ is also a regular singular point. Therefore, Eq. (S5) can be transformed into Gauss hypergeometric Eq. (GHE) \cite{slavianov2000special}. \\
  By considering $qz=\Tilde{z}$ we can convert Eq. (S5) into
  
\begin{align}
 \Tilde{z}(1-\Tilde{z})\derivative[2]{G(\Tilde{z})}{\Tilde{z}}-\xi G(\Tilde{z})=0 
 \end{align}
  
  Putting $G(\Tilde{z})=C_1\hspace{1mm} \Tilde{z}w(\Tilde{z})$ as an ansatz solution in Eq. (S6), it can be written as \\
 \begin{align}
 \Tilde{z}(1-\Tilde{z})\derivative[2]{w(\Tilde{z})}{\Tilde{z}} + 2 (1-\Tilde{z}) \derivative{w(\Tilde{z})}{\Tilde{z}} - \xi w(\Tilde{z}) =0.
 \end{align}
  
 Note that Eq. (S7) is of the form of GHE \cite{abramowitz1988handbook},\\
 \begin{align}
 x(1-x)\derivative[2]{W(x)}{x} + \left[c - (a +b + 1)x\right]\derivative{W(x)}{x} - a b W(x)=0
 \end{align}
 with $a=\frac{1}{2}(1+\sqrt{1-4\xi)}$, $b=\frac{1}{2}(1-\sqrt{1-4\xi)}$ and $c=2$. Therefore, $G(z)=C\hspace{1mm} qz \hspace{1mm}{_2F_1\left(a,b;c;qz\right)}$  is the only independent solution of Eq. (S7) since $c$ is a positive integer. Constant, $C$ in $G(z)$ can be determined using the boundary condition $G(1)=1$. Therefore, we can calculate $G(z)$ as
 \begin{align}
    G(z)= \frac{ z \hspace{1mm}{_2F_1\left(a,b;c;qz\right)}}{\hspace{1mm}{_2F_1\left(a,b;c;q\right)}} 
 \end{align}
 and the probability distribution  as \\
 \begin{align}
     P(n)= \frac{1}{_2F_1\left(a,b;c;q\right)} \frac{(a)_{n-1}(b)_{n-1}}{(c)_{n-1}} \frac{q^{n-1}}{(n-1)!}
 \end{align}
  where $(a)_n=a (a+1) (a+2)...(a+n-1) $ is the Pochhamer symbol.


\section{Bursty Synthesis-Fission-Decay}

 The generating function's equation is \\
  \begin{align}
 (1-\eta z)\derivative{G(z)}{z}-\left[\frac{a\eta q}{1-zq}\right]G(z)=0 
  \end{align}
 where $\alpha=k^0_d/k_{fis}$ and $\eta=k_{fis}/\gamma$. Integrating over z \\
 
 \begin{align*}
 \int \frac{dG}{G}=\frac{\alpha \eta q}{\eta - q}\int\left[\frac{\eta}{1-\eta z}-\frac{q}{1-zq}\right]dz
 \end{align*}
 
 and applying the boundary condition, $G(z)$ can be calculated as\\
 \begin{align}
     G(z)=\left[\frac{(1-qz)(1-\xi)}{(1-\eta z)(1-q)}\right]^{\left(\frac{\alpha \eta q}{\eta - q}\right)} = \left[\frac{1-\eta}{1-q}\right]^m \left[\frac{1-qz}{1-\eta z}\right]^m
\end{align}
where $m=\left(\frac{\alpha \eta q}{\eta - q}\right)$.

 Using binomial expansion  Eq. (S12) can be written as\\
 \begin{align}
 \left[\frac{1-\eta}{1-q}\right]^m(1-qz)^m(1-\eta z)^{-m}=\left[\frac{1-\eta}{1-q}\right]^m\sum_{k=0}^{\infty}(-1)^k\binom{m}{k}(qz)^k \sum_{k=0}^{\infty}\binom{k+m-1}{k}(\eta z)^k.
 \end{align}

Note that the binomial coefficient for real arguments can be expressed as $\binom{m}{k}=\left[\frac{\Gamma(m+1)}{\Gamma(k+1)\Gamma(1+m-k))}\right]$.
 Next using the Cauchy product, given by
\begin{align}
\sum_{n=0}^{\infty}a_{n}x^n\sum_{n=0}^{\infty}b_{n}x^n=\sum_{n=0}^{\infty}\left(\sum_{k=0}^{n}a_{k}b_{n-k}\right)x^n ,
\end{align}

 and Eq. (S13) we calculated the $n^{\textit{th}}$ term of $G(z)$, $P(n)$  as\\
 \begin{align*}
 &\left[\frac{1-\eta}{1-q}\right]^m \sum_{k=0}^{n}(-1)^k\binom{m}{k}\binom{n-k+m-1}{n-k} q^k \eta^{n-k} \nonumber \\ & = \left[\frac{1-\eta}{1-q}\right]^m \eta^{n}\sum_{k=0}^{n}(-1)^k \left(\frac{q}{\eta}\right)^{k} \left[\frac{\Gamma(m+1)}{\Gamma(k+1)\Gamma(1+m-k)}\right] \left[\frac{\Gamma(n-k+m)}{\Gamma(n-k+1)\Gamma(m)}\right] \nonumber \\ & = \left[\frac{1-\eta}{1-q}\right]^m \frac{\Gamma(n+m)}{\Gamma(m)n!} \eta^{n}\sum_{k=0}^{n} \left(\frac{q}{\eta}\right)^{k}\frac{1}{k!}\left[(-1)^k\frac{\Gamma(m+1)}{\Gamma(1+m-k)}\right]\left[(-1)^k\frac{\Gamma(n+m-k)}{\Gamma(n+m)}\right]\left[(-1)^k\frac{n!}{(n-k)!}\right] \nonumber \\ & = \left[\frac{1-\eta}{1-q}\right]^m \frac{\Gamma(n+m)}{\Gamma(m)n!}\eta^{n}\sum_{k=0}^{n}(-1)^k \frac{1}{k!}\left(\frac{q}{\eta}\right)^{k} \left[\frac{(-n)_{k}(-m)_{k}}{(1-n-m)_{k}}\right]\nonumber \\ & =\left[\frac{1-\eta}{1-q}\right]^m \frac{\Gamma(n+m)}{\Gamma(m)n!}\eta^{n} \hspace{1mm}{_2F_1\left(-n,-m;1-n-m;\frac{q}{\eta}\right)}.
 \end{align*}
 
 Here we used the expression of Pochhamer symbol for negative argument $(s)_n=\frac{(-1)^n\Gamma(1-s)}{\Gamma(1-s-n)}$ where $s\in -\mathbb{R}$.



 \section{Bursty Synthesis-Fusion-Decay}
 
 For this submodel, the equation of generating function is \\
 \begin{align}
   z(1-qz)\derivative[2]{G(z)}{z} + \delta(1 - qz) \derivative{G(z)}{z} - \xi qG(z)=0  
 \end{align}
 where $\delta=\frac{\gamma}{k_{fus}}$ and $\xi=\frac{k^0_d}{k_{fus}}$. Here all singular points, $z=0,1/q$ and $\infty$ are regular. Hence Eq. (S15) can be expressed in the form of GHE, Eq. (S8). By doing variable transformation, $qz=\Tilde{z}$, Eq. (S15) can be expressed as \\
 \begin{align}
   \Tilde{z}(1-\Tilde{z})\derivative[2]{G(\Tilde{z})}{\Tilde{z}} + \delta(1 - \Tilde{z}) \derivative{G(\Tilde{z})}{\Tilde{z}} - \xi G(\Tilde{z})=0  
 \end{align}
 Comparing Eq. (S8) and (S16) we can write the solution as $G(z)=C\hspace{1mm} {_2F_1\left(a',b';c';qz\right)}$ with $a'=\frac{1}{2}\left[\delta - 1 +  \sqrt{(\delta - 1)^2-4\xi}\right]$, $b'=\frac{1}{2}\left[\delta - 1 - \sqrt{(\delta - 1)^2-4\xi)}\right]$ and $c'=\delta$. The other independent solution is discarded as $c'=\delta$ can be a positive integer. Note that depending on the value of parameters $\delta$ and $\xi$, $a'$ and $b'$ can be complex conjugate to each other. Hence their product will always be a real number. Following a similar procedure as the previous cases, we calculated the probability distribution as \\
 \begin{align}
     P(n)= \frac{1}{_2F_1\left(a',b';c';q\right)} \frac{(a')_{n}(b')_{n}}{(c')_{n}} \frac{q^{n}}{n!} \nonumber
 \end{align}


\subsection{Fusion reduces noise: noise decomposition}

 The Master equation for the `Bursty synthesis-Fusion-Decay' submodel is \\
\begin{align}
    \frac{d P(n,t)}{dt} = k^0_d\sum_{k=0}^{n-1} (1-q){q}^{n-k} P(k,t) + d_{n+1} P(n+1,t) - (k^0_d q + d_n) P(n,t)
\end{align}
where $d_n = \gamma n+k_{fus}n(n-1)$.

To calculate the equation of mean, we multiply Eq. (S17) by $n$ and sum over all $n$. First, consider the terms of the RHS.

\underline{1st term:}\\
 
  $k^0_d\sum_{k=0}^{n-1} (1-q){q}^{n-k} P(k)$\\
 
 $=k^0_d(1-q)\sum_{i=0}^{\infty}\sum_{j=1}^{\infty} q^j P(i) \delta_{i+j,n}$ \hspace{1.0cm}($\delta_{p,q}=1$, for $p=q$, and $0$ otherwise)\\
 
 Multiplying by $n$ and summing over $n$\\

 $k^0_d(1-q)\sum_{n=0}^{\infty}n\sum_{i=0}^{\infty}\sum_{j=1}^{\infty} q^j P(i) \delta_{i+j,n}$ \hspace{0.5cm}\\

$=k^0_d(1-q)\sum_{j=1}^{\infty}\sum_{i=0}^{\infty} (i+j) q^j P(i) $\\

$=k^0_d(1-q)\left[ \sum_{i=0}^{\infty} iP(i)\sum_{j=1}^{\infty}q^j + \sum_{i=0}^{\infty}P(i) )\sum_{j=1}^{\infty} j q^j  \right]$ \\

$=k^0_d \left[ q\langle n \rangle + \frac{q}{1-q} \right]$\\

\underline{2nd term:}\\

$d_{n+1} P(n+1)$\\

Multiplying by $n$ and summing over $n$\\

$\sum_{n=0}^{\infty}n d_{n+1} P(n+1)$\\

$=\sum_{m=1}^{\infty} (m-1)d_m P(m)$  \hspace{0.5cm} (Considering, $n+1\equiv m$)\\

$=\langle nd_n \rangle - \langle d_n \rangle$\\

\underline{3rd term:}\\

$k^0_d q P(n)$\\

Multiplying by $n$ and summing over $n$\\

$k^0_d q \sum_{n=0}^{\infty} nP(n) = k^0_d q \langle n \rangle$\\

\underline{4th term:}\\

$d_n P(n)$\\

Multiplying by $n$ and summing over $n$\\

$\sum_{n=0}^{\infty}n d_n P(n) = \langle nd_n \rangle$\\

Therefore, by adding all the terms, the equation of mean is 

\begin{align}
    \frac{d\langle n \rangle}{dt}  =  k^0_d B - \langle d_n \rangle 
\end{align} 
where $B= \frac{q}{1-q}.$

Similarly, it is easy to check that
\begin{align*}
    \frac{d \langle n^2 \rangle}{dt}=2k^0_d B\langle n \rangle + k^0_d B(1+2B) -2 \langle nd_n \rangle + \langle d_n \rangle.
\end{align*}

Hence, the equation of variance is 
\begin{align}
    \frac{d\sigma^2}{dt} &= \frac{d\langle n^2 \rangle}{dt} - 2\langle n \rangle \frac{d\langle n \rangle}{dt} \nonumber \\ & = k^0_d B(1+2B) -2 Cov (n,d_n) + \langle d_n \rangle,
\end{align}
where $ Cov (n,d_n)= \langle nd_n \rangle - \langle n \rangle \langle d_n \rangle $ is the covariance between $n$ and $d_n$. Note that at the steady state, Eq. (S18) and (S19) do not explicitly have $\langle n \rangle$ and $\sigma^2$ terms as they are embedded in $d_n$. In detail $\langle d_n \rangle = \gamma \langle n \rangle + k_{fus} \left(\langle n^2 \rangle - \langle n \rangle \right) $ and $ Cov (n,d_n)= \sigma^2 (\gamma - k_{fus})+ k_{fus}\left( \langle n^3 \rangle - \langle n \rangle \langle n^2 \rangle\right)$. Moreover, due to fusion, the moment equations are not closed and contain a hierarchy. Therefore, to compute noise decomposition we will compare the effect of fusion in noise level over the known `Bursty synthesis-decay' case. 

\par
Consider $d_n=\gamma n+f_n$ where $f_n$ is the fusion propensity (here, $k_{fus}n(n-1))$. Consequently, $\langle d_n \rangle = \gamma \langle n \rangle +\langle f_n \rangle $ and $Cov (n,d_n)= \gamma \sigma^2 + Cov (n,f_n) $. Using Eq. (S18) and (S19), at the steady state we obtain
\begin{align*}
    \langle n \rangle=\frac{1}{\gamma}(k^0_d B - \langle f_n \rangle), \hspace{3mm}\sigma^2=\frac{1}{\gamma}(k^0_d B(1+B) - Cov (n,f_n)).
\end{align*}

Note that $\langle n \rangle$ and $\sigma^2$ are positive real numbers. Since $n$ is a positive integer including $0$, $Cov(n,f_n) > 0$ \cite{jia2017stochastic}, fusion decreases the mean and variance over `Bursty synthesis-decay' limit ( in this case mean and variance are $k^0_d B/\gamma$ and $\left(k^0_d B(1+B)\right)/\gamma$ respectively) by  $\langle f_n\rangle/\gamma$ and $Cov(n,f_n)/\gamma$ respectively. Therefore, the Fano factor ($FF=\sigma^2/\langle n \rangle$) is
\begin{align}
    FF = FF_{\text{syn-decay}}-\Delta FF = \frac{(k^0_d B(1+B) - Cov (n,f_n))}{(k^0_d B - \langle f_n \rangle)}
\end{align}
where  $FF_{\text{syn-decay}}=1+B$ is the Fano factor of the `bursty synthesis-decay' submodel, and $\Delta FF =\frac{Cov(n,f_n)-(1+B)\langle f_n \rangle}{k^0_d B-\langle f_n \rangle}$ is the noise contributed by the fusion process. It can be shown that $\Delta FF$ is positive. Using the analytical distribution function for the synthesis-fusion-decay model, we verified this by plotting $\Delta FF$  as a function of $k_d^0$ for both high and low limits of the other parameters, $k_{fus}$, $\gamma$, and $q$ (see Fig. S2). Therefore, the inclusion of the fusion process reduces the noise compared to the `Bursty synthesis-decay', and hence fusion may be considered as negative feedback.

  \section{Derivation of Generating Function's Equation for the full model}

 Eq (1). in the steady state can be written as
 \begin{align}
    &k^0_d\sum_{n^{'}=0}^{n-1} (1-q){q}^{n-n^{'}} P(n^{'})+ \gamma(n+1)P(n+1)  + k_{fis}(n-1)P(n-1) + k_{fus}n(n+1)P(n+1) \nonumber \\ & -\left [k^0_d\sum_{n^{'}=n+1}^{\infty} (1-q){q}^{n^{'}-n} + \gamma n + k_{fis}n + k_{fus}n(n-1)\right ]P(n)=0
  \end{align}
  To formulate Eq. (14) in terms of $G(z)\left(=\sum_{n=0}^{\infty} z^n P(n)\right)$, we first multiply each term of Eq. (S22) by $z^n$, and then we sum over all $n$.

 \underline{1st term:}\\

 $k^0_d\sum_{n^{'}=0}^{n-1} (1-q){q}^{n-n^{'}} P(n^{'})$\\
 
 $=k^0_d(1-q)\sum_{i=0}^{\infty}\sum_{j=1}^{\infty} q^j P(i) \delta_{i+j,n}$ \hspace{1.0cm}($\delta_{p,q}=1$, for $p=q$, and $0$ otherwise)\\
 
 Multiplying by $z^n$ and summing over $n$\\
 
 $k^0_d(1-q)\sum_{n=0}^{\infty}z^n\sum_{i=0}^{\infty}\sum_{j=1}^{\infty} q^j P(i) \delta_{i+j,n}$ \hspace{0.5cm}\\
 
 $=k^0_d(1-q)\sum_{j=1}^{\infty}z^jq^j\sum_{i=0}^{\infty} z^i P(i) $\\
 
 $=k^0_d(1-q)\sum_{j=1}^{\infty}z^jq^j G(z) $\\

 $=\frac{k^0_dq(1-q)G(z)z}{1-zq} $ \hspace{0.5cm}  (Since, $|zq| < 1$,  using infinite GP series formula)\\

 \underline{2nd term:}\\

 $\gamma(n+1)P(n+1)$\\
 
 Multiplying by $z^n$ and summing over $n$\\
 
 $\gamma\sum_{n=0}^{\infty}z^n(n+1)P(n+1)=\gamma \frac{dG}{dz}$\\

\underline{3rd term:}\\

$k_{fis}(n-1)P(n-1)$ \\

Multiplying by $z^n$ and summing over $n$\\
 
$k_{fis}\sum_{n=0}^{\infty}z^n(n-1)P(n-1)$  \\ 

$=k_{fis}\sum_{k=0}^{\infty}z^{k+1}k P(k)$ \hspace{0.5cm} (Considering, $n-1\equiv k$)\\

$= k_{fis}z^2 \frac{dG}{dz}$\\

 \underline{4th term:}\\ 
 
 $k_{fus}n(n+1)P(n+1)$\\

 Multiplying by $z^n$ and summing over $n$\\
 
 $k_{fus}\sum_{n=0}^{\infty}z^n n(n+1)P(n+1)$  \\

 $=k_{fus}\sum_{k=0}^{\infty}z^{k-1} k(k-1)P(k)$ \hspace{0.5cm} (Considering, $n+1 \equiv k$)\\
 
 $= k_{fus}z\derivative[2]{G(z)}{z} $\\
 
 \underline{5th term:}\\ 
 
 $k^0_d\sum_{n^{'}=n+1}^{\infty} (1-q){q}^{n^{'}-n}P(n)=k^0_d(1-q)\sum_{j=1}^{\infty} {q}^jP(n)$\\
 
 $=k^0_dqP(n)$\\
 
 Multiplying by $z^n$ and summing over $n$\\
 
 $k^0_dq\sum_{n=0}^{\infty}z^n P(n)=k^0_dq G(z)$\\

 \underline{6th term:}\\
 
 $\gamma n P(n)$\\
 
 Multiplying by $z^n$ and summing over $n$\\
 
 $\gamma\sum_{n=0}^{\infty}z^n n P(n)=\gamma z \frac{dG}{dz}$\\

 \underline{7th term:}\\
 
 $k_{fis} n P(n)$\\
 
 Multiplying by $z^n$ and summing over $n$\\
 
 $k_{fis}\sum_{n=0}^{\infty}z^n n P(n)=k_{fis} z \frac{dG}{dz}$\\

 \underline{8th term:}\\
 
 $k_{fus}n(n-1)P(n)$\\
 
 Multiplying by $z^n$ and summing over $n$\\
 
 $k_{fus}\sum_{n=0}^{\infty}z^n n(n-1)P(n)$\\
 
 $=k_{fus} z^2 \derivative[2]{G(z)}{z} $\\ 
 
 By adding all the above terms we obtain the generating function's equation for the full model,
\begin{align}
 z\derivative[2]{G(z)}{z} + (\delta - \phi z) \derivative{G(z)}{z} - \alpha \phi \frac{q}{1-qz}G(z)=0  ,
 \end{align}
 where $\alpha=k^0_d/k_{fis},  \delta=\gamma/k_{fus}$, and $\phi=k_{fis}/k_{fus}$.

 
 \section{Singularity analysis of generating function's equation for the full model}

Consider a linear, homogeneous, second-order differential equation of the form\\
\begin{align}
    \derivative[2]{G(z)}{z} + p(z) \derivative{G(z)}{z} - q(z)G(z)=0
\end{align}
Here, $p(z)=\frac{\delta - \phi z}{z}$ and $q(z)=-\frac{\alpha \phi q}{z(1-qz)}$ (see Eq. (14) or (S23) ).
 Therefore singular points are at $z=0, 1/q$, and $\infty $. \\
 
 At $z=0$,\\
 \[\lim_{z\to 0} zp(z)=\delta\] and \[\lim_{z\to 0} z^2q(z)=\lim_{z\to 0}\frac{-\alpha \phi q}{z(1-qz)}=0\]\\
 both exist and are finite. Hence $z=0$ is a regular singular point.\\
 
 Similarly, at $z=\frac{1}{q}$,\\
 \[\lim_{z\to\frac{1}{q}} (z-1/q)p(z)=\lim_{z\to\frac{1}{q}}\frac{(qz-1)(\delta - \phi z)}{qz}=0\] and  \[\lim_{z\to\frac{1}{q}} (z-1/q)^2q(z)=\lim_{z\to\frac{1}{q}} \frac{\alpha \phi (qz-1)}{qz}=0\]\\
 
 both exist and are finite. Therefore, $z=1/q$ is a regular singular point.\\
 
 \subsection{Singularity at Infinity}
 
To check singularity at $\infty$, consider $t=1/z$. Therefore,
 $\frac{d}{dz} = -t^2\frac{d}{dt}$, and
 $\derivative[2]{}{z}=t^4\derivative[2]{}{t}+2t^3\derivative{}{t} $.\\
 
 Now Eq. (S24) can be written in terms of the variable $t$ as, \\
 
 \begin{align}
     t^4 \derivative[2]{G}{t}+2t^3\derivative{G}{t}+p(\frac{1}{t})(-t^2\derivative{G}{t})+q(\frac{1}{t})G(\frac{1}{t})=0 \nonumber
 \end{align}
 \begin{align}
 \implies  \derivative[2]{G}{t}+\frac{2t-p(1/t)}{t^2}\derivative{G}{t}+\frac{q(1/t)}{t^4}G(1/t)=0  
\end{align}

Now it is equivalent to check the singularity of Eq. (S25) at $t=0$ and the singularity of Eq. (S24) at $z=\infty$.
\par
 Here, we can write $p(1/t)\equiv p(t)=(\delta-\frac{\phi}{t})t=(\delta t-\phi)$, and $q(1/t)\equiv q(t)=-\frac{\alpha \phi q}{1/t(1-q/t)}=t^2\left[\frac{\alpha \phi q}{q-t}\right]$\\

Therefore,

\[\lim_{t\to 0} t\frac{2t-p(1/t)}{t^2}=\lim_{t\to 0}\frac{(2-\delta)t+\phi}{t}=\infty\]  and \[\lim_{t\to 0} t^2\frac{q(1/t)}{t^4}=\lim_{t\to 0}\frac{\alpha \phi q}{(q-t)}=\frac{\alpha \phi q}{q}\]\\

Hence, $t=0$ or $z=\infty$ is an irregular singular point.\\

 Therefore, $z=0, 1/q$ are regular singular points and $z=\infty $ is an irregular singular point. So  Eq. (14) (or (S23)) is not of Riemann type and can not be converted to GHE.


 \section{Full model: Bursty Synthesis-Fission-Fusion-Decay}
 
 Considering only the linear coefficients in $z$ we approximated Eq. (14) as 
 \begin{align}
 z\derivative[2]{G(z)}{z} + (\delta - \phi z) \derivative{G(z)}{z} - \alpha \phi q(1+qz)G(z)=0,
 \end{align}

 where $\alpha=\frac{k^0_d}{k_{fis}},  \delta=\frac{\gamma}{k_{fus}}$, and $\phi=\frac{k_{fis}}{k_{fus}}$  are all real positive constants.

 Since it has linear coefficients in $z$ it can be transformed into the confluent hypergeometric differential equation (Kummer's equation) \cite{slavianov2000special}. Considering an ansatz $G(z)=C \hspace{1mm}\exp{\lambda z}w(z)$ and plugging it in Eq. (S26) we can obtain ,
 
 \begin{align}
 z\derivative[2]{w(z)}{z} + \left[\delta + (2\lambda-\phi) z\right] \derivative{w(z)}{z} + \left[z(\lambda^2-\phi \lambda -A_1)-\Tilde{\alpha}+\delta \lambda\right]w(z)=0
 \end{align}
 
 where $\Tilde{\alpha}=\alpha\phi q$ and $A_1=\alpha\phi q^2$. To reduce Eq. (S27) to Kummer's equation we choose $\lambda=\frac{1}{2}\left(\phi-\sqrt{\phi^2+4A_1}\right)$ and $\tilde z=z\sqrt{\phi^2+4A_1}$ and substitute it in (S27). Finally, it converts to 
 
 \begin{align}
 \tilde z\derivative[2]{w(\tilde z)}{\tilde z} + \left(\delta - \tilde z\right) \derivative{w(\tilde z)}{\tilde z} - \left[\frac{\Tilde{\alpha}-\frac{\delta}{2}\left(\phi-\sqrt{\phi^2+4A_1}\right)}{\sqrt{\phi^2+4A_1}}\right]w(\tilde z)=0.
 \end{align}
 
 Therefore, $w(\tilde z)=C'\hspace{1mm}  _1F_1\left(\left[\frac{\Tilde{\alpha}-\frac{\delta}{2}\left(\phi-\sqrt{\phi^2+4A_1}\right)}{\sqrt{\phi^2+4A_1}}\right],\delta;\tilde z\right)$ and \\
 
 $G(z)=C\exp{\frac{z}{2}\left(\phi-\sqrt{\phi^2+4A_1}\right)}{_1F_1\left(\left[\frac{\Tilde{\alpha}-\frac{\delta}{2}\left(\phi-\sqrt{\phi^2+4A_1}\right)}{\sqrt{\phi^2+4A_1}}\right],\delta;z\sqrt{\phi^2+4A_1}\right)}$\\
 
 For convenience we write $G(z)=C \exp{\frac{\mu z}{2}}{_1F_1\left(\kappa,d;\zeta z\right)}$, 
 where $\zeta=\sqrt{\phi^2+4A_1}$, $d=\delta$, $\mu=\left(\phi-\zeta\right)$, and $\kappa=\left(\frac{\Tilde{\alpha}-\frac{d}{2}\mu}{\zeta}\right)$.  \\

 Imposing boundary condition $G(1)=1$ yields ,
 \begin{align*}
 C=\frac{1}{\exp{\frac{\mu }{2}}{_1F_1\left(\kappa,d;\zeta \right)}}.
 \end{align*}
 To calculate the $P(n)$ we use the Cauchy Product (Eq. S14). Utilising it, the $n$-th term of $G(z)$ is obtained as
 \begin{align}
 P(n)=C \sum_{k=0}^{n}\frac{(\kappa)_k}{(d)_k}\frac{\zeta^k}{k!} \frac{(\mu/2)^{n-k}}{(n-k)!}= C \left({\frac{\mu}{2}}\right)^n\frac{_2F_1\left(-n,\kappa;d;-\frac{2\zeta}{\mu}\right)}{n!} .   
 \end{align}\\
 
\section{Supplemental figures}

\begin{figure*}[hbt!]
\begin{center}
{\includegraphics[width=0.9\textwidth]{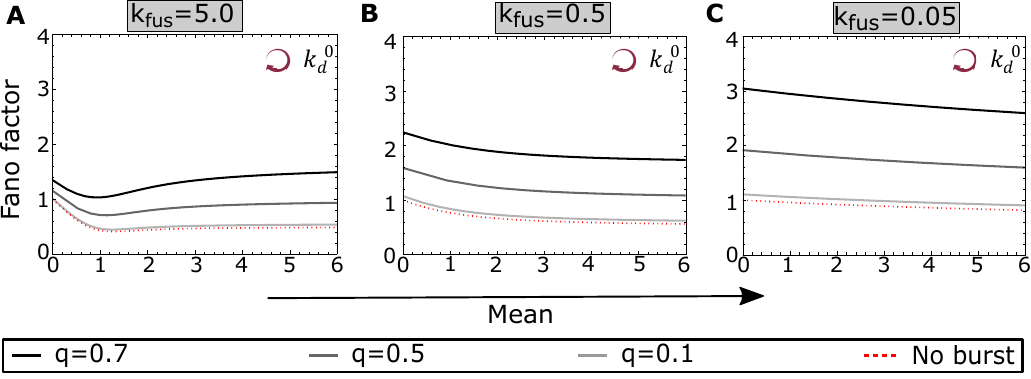}
\caption{ {\bf Decreasing the fusion rate reduces the nonmonotonicity in the Fano-factor-versus-mean curves of the `Bursty synthesis-fusion-decay' submodel, implying a transit toward `Bursty synthesis-decay' model.} (A-C) We tuned $k^0_d$ for the `synthesis-fusion-decay' submodel and kept other rates fixed ($\gamma=1$ for A-C and $k_{fus}=5, 0.5$, and $0.05$ for A, B and C, respectively). Comparing A-C, it is evident that the dip near mean $\sim$ 1 in the noise profile (Panel A) results from fusion dominance over decay. 
When $k_{fus} \ll \gamma$ (Panel C), the noise profile looks like `synthesis-decay' with Fano factor $\sim 1/(1-q)$ (compare with Fig 2A in the main text). All rate parameters are in arbitrary units of time$^{-1}$. In all panels, shades of grey represent different $q$ values, and red dashed curves represent the `no-burst' cases.   }}\label{figS1}
\end{center}
\end{figure*}


\begin{figure*}[hbt!]
\begin{center}
{\includegraphics[width=0.9\textwidth]{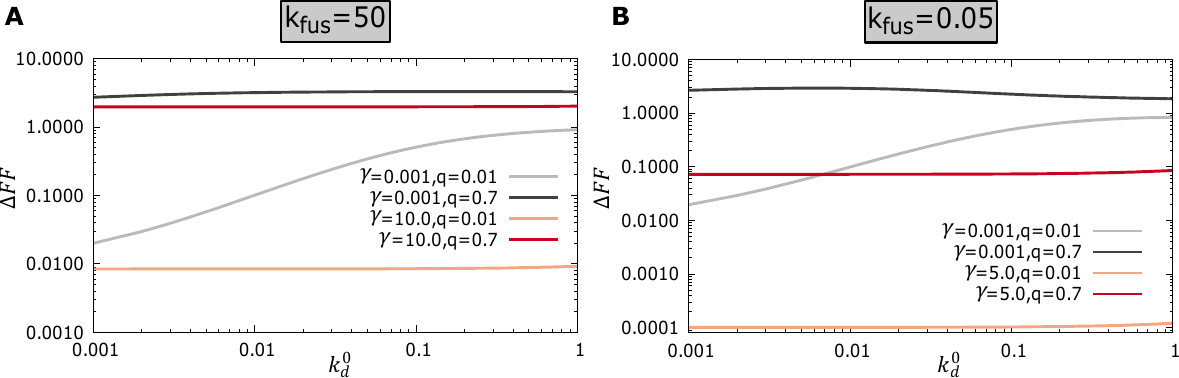}
\caption{ {\bf Fusion always suppresses noise compared to the `Bursty synthesis-decay' submodel.} (A-B) We plotted $\Delta FF$, the noise contributed by the fusion process (see Eq. S20), with $k_d^0$ for high and low values of $k_{fus}$. To verify that $\Delta FF$ is always positive, we checked both the high and low limits of the parameters $\gamma$ and $q$. Different color shades represent those cases. }}
\end{center}
\end{figure*}

 \begin{figure*}[hbt!]
\begin{center}
{\includegraphics[width=0.9\textwidth]{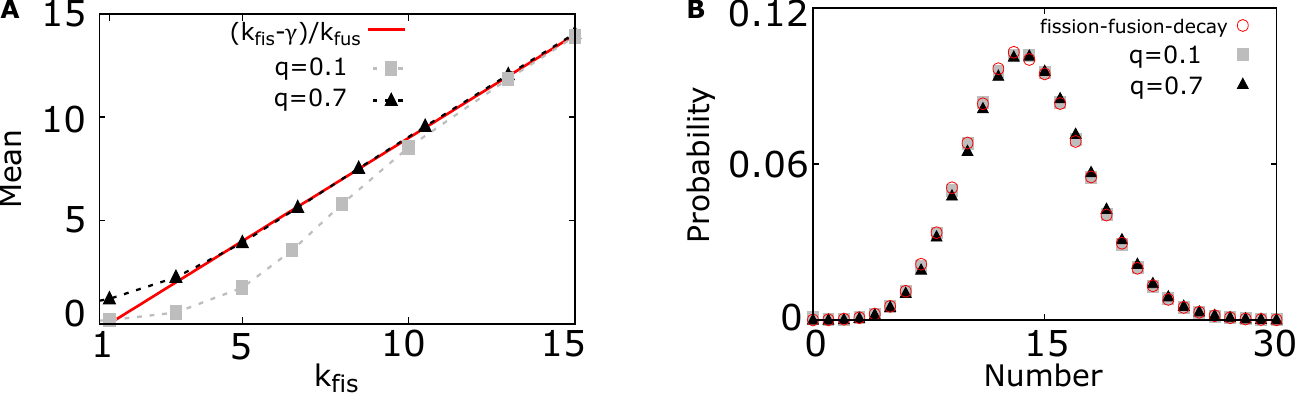}
\caption{ {\bf At a high fission rate ($k_{fis}$), the full model boils down to the `fission-fusion-decay' sub-model.} (A) The mean organelle number is shown as a function of $k_{fis}$ for the `fission-fusion-decay' sub-model and the full model for different values of $q$. The red straight line is the steady-state solution of the mean-field equation for the `fission-fusion-decay' submodel (the mean-field equation is given by $\dot{n}=k_{fis}n-\gamma n-k_{fus}n^2$). Different shades of grey symbols and dashed lines represent the mean organelle number for the full model obtained from exact stochastic simulations. Note that for higher fission rates ($k_{fis}>10$), the exact means of the full model match well with the steady-state mean-field solution of the `fission-fusion-decay' sub-model, irrespective of the values of $q$. (B) Organelle number distributions are shown at steady-state for the `fission-fusion-decay' sub-model and the full model. At a high fission rate ($k_{fis}=15$), probability distributions of the `fission-fusion-decay' submodel (red circles) and the full model (grey squares and black triangles for different $q$ values) overlap together onto a bell-shaped curve. These distributions are obtained by stochastic simulations. We used $\gamma=k_{fus}=1$ for both panels. All rate constants are in arbitrary units of time$^{-1}$.  }}\label{figS3}
\end{center}
\end{figure*}

\begin{figure*}[hbt!]
\begin{center}
{\includegraphics[width=0.9\textwidth]{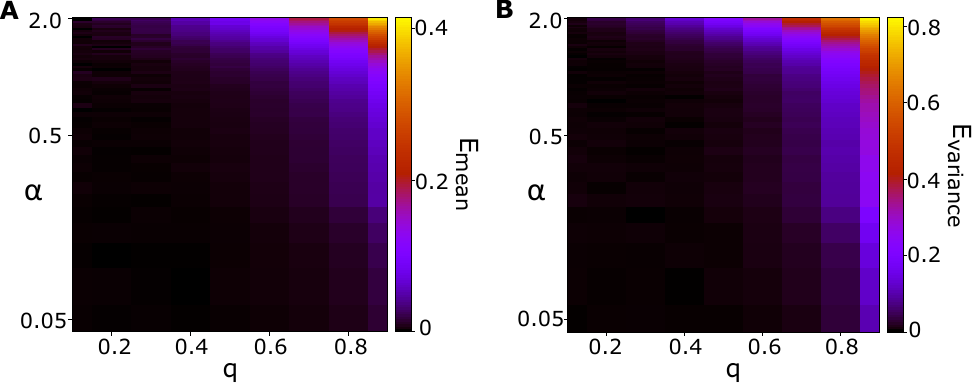}
\caption{ {\bf Relative errors in mean and variance for the full model.} (A) Heatmap shows the intensity of relative error in mean calculated using the equation: $E_{mean}=|\langle n\rangle_{Gillespie}-\langle n\rangle_{analytics}|/\langle n\rangle_{Gillespie}$,  where $\langle n \rangle_{Gillespie}$ is the mean obtained from exact Gillespie simulations and $\langle n\rangle_{analytics}$ is the approximate mean calculated using Eq. 18. (B) Similarly, the heatmap shows the error in variance calculated using $E_{variance}=|\sigma^2_{Gillespie}-\sigma^2_{analytics}|/\sigma^2_{Gillespie}$, where  $\sigma^2_{Gillespie}$ is the variance from exact simulations and $\sigma^2_{analytics}$ is calculated from Eq. 19. We kept  $k^0_d=k_{fus}=\gamma=1$ for all panels.  }} \label{figS4}
\end{center}
\end{figure*}


\end{document}